# Optomechanical and Photothermal Interactions in Suspended Photonic Crystal Membranes

David N. Woolf[1], Pui-Chuen Hui[1], Eiji Iwase[1,3], Mughees Khan[1,4], Alejandro W. Rodriguez[1,2], Parag Deotare[1,5], Irfan Bulu[1], Steven G. Johnson[2], Federico Capasso[1], Marko Loncar[1*]

[1]School of Engineering and Applied Sciences, Harvard University, Cambridge, MA 02138 USA
[2]Department of Mathematics, Massachusetts Institute of Technology, Cambridge, MA 02139 USA
[3]Department of Applied Mechanics and Aerospace Engineering, Waseda University, Tokyo, Japan 169-8050
[4]Wyss Institute, Harvard University, Boston, MA 02115
[5]Department of Electrical Engineering, Massachusetts Institute of Technology, Cambridge, MA 02139 USA
*loncar@seas harvard.edu

**ABSTRACT:**

**We present here an optomechanical system fabricated with novel stress management techniques that allow us to suspend an ultrathin defect-free silicon photonic-crystal membrane above a Silicon-on-Insulator (SOI) substrate with a gap that is tunable to below 200 nm. Our devices are able to generate strong attractive and repulsive optical forces over a large surface area with simple in- and out-coupling and feature the strongest repulsive optomechanical coupling in any geometry to date ($g_{0M}/2\pi \approx$ -65 GHz/nm). The interplay between the optomechanical and photo-thermal-mechanical dynamics is explored, and the latter is used to achieve cooling and amplification of the mechanical mode, demonstrating that our platform is well-suited for applications in low-power mass, force, and refractive-index sensing as well as and optomechanical accelerometry.**

Devices exhibiting resonant mechanical dynamics have applications ranging from high-precision mass and force sensing in nanoelectromechanical systems (NEMS)[1-5] to novel quantum manipulation enabled by ground-state cooling of sub-micron-scale mechanical objects.[6-11] Additionally, the concentration of light into small volumes has been shown to have broad applications due to the sensitivity of the optical mode properties to its local environment.[12,13] Recently, there have been rapid developments in the field of optomechanics that utilize light to actuate a new class of low-mass compact resonators[3,13-16] and that push the limits of device scalability.[1,5,17-19] In particular, optomechanical devices that can be fully integrated onto a silicon chip can act as active sensors or reconfigurable elements in chip-based systems.[15,19-21] Here, we present a versatile optomechanical structure fabricated with novel stress management techniques that allow us to suspend an ultrathin defect-free silicon photonic-crystal membrane above a Silicon-on-Insulator (SOI) substrate with a gap that is tunable to below 200 nm. Our devices are able to generate strong attractive and repulsive optical forces over a large surface area and feature the strongest repulsive optomechanical coupling in any geometry to date ($g_{OM}/2\pi \approx -65$ GHz/nm). The interplay between the optomechanical and photo-thermal-mechanical dynamics is explored, and the latter is used to achieve cooling of the mechanical mode from room temperature down to 5 K with approximately one milliwatt of power and amplification of the mode to achieve three orders of magnitude of linewidth narrowing and oscillation amplitudes of approximately 1 nm. We achieve these figures by leveraging a delocalized "dark" mode of our optical system that minimize the impact of two-photon absorption while simultaneously generating large forces in our devices. Owing to the simplicity of the in- and out-coupling of light as well as its large surface area, our platform is well-suited for applications in both mass sensing (with sub-femtogram resolution), and refractive index sensing ($\delta\lambda/\delta n_{eff} \approx 100$ nm per unit refractive index) and optomechanical accelerometry.

It has been previously shown[22-24] that two co-propagating modes at optical frequency $\omega$ in parallel dielectric waveguides separated by a distance s interact evanescently, resulting in "bonding" and "antibonding" eigenmodes of the structure. As s decreases, the coupling between the two waveguides increases, decreasing the eigenfrequency of the bonding mode and increasing the eigenfrequency of the antibonding mode. The force between the two waveguides can be written as $F_{opt} = U_{ph}g_{OM}/\omega$, where $U_{ph} = N\hbar\omega$ is the energy flowing in the waveguides, N is the number of photons in the mode, and $g_{OM}$ is the optomechanical (OM) coupling coefficient, defined as $d\omega/ds$. In the "bonding" configuration, the fields in the two waveguides are in-phase and generate an attractive force ($d\omega/ds > 0$), while the "antibonding" configuration corresponds to out-of-phase fields and a repulsive ($d\omega/ds < 0$) interaction. Because the above expression for the force is general,[25] we can extend it beyond the parallel waveguide geometry and apply it to our platform, shown in Figure 1(a).

The platform consists of a square silicon photonic crystal (PhC) slab containing a 30 × 30 array of holes with periodicity p = 0.92μm and hole diameter d = 0.414μm, as defined in the illustration in Fig. 1(b), suspended a few hundred nanometers above a Silicon-on-Insulator (SOI) substrate and is capable of generating strong attractive and repulsive forces.[26] To fabricate our devices, we oxidize the top 35 nm of two SOI wafers (device layer thickness = 220 nm, buried oxide layer thickness = 2 μm) and bond the two oxidized surfaces together. After removal of one of the handle wafers along with its buried oxide layer, we are left with a double-device layer SOI with two thin silicon layers of thickness h = 185 nm separated by a thin oxide layer of thickness $s_0$ = 260 nm. Using e-beam lithography, we write our device pattern into a thin layer of electron-beam resist, which then acts as a mask as we etch through the top silicon layer using Reactive Ion Etching. The top membrane is then released by removing the thin oxide layer with Hydrogen Fluoride

Vapor-phase Etching (HFVE) (see Methods/Supplement for details). The device parameters *p* and *d* were chosen to result in an antibonding mode (profile shown in Fig. 1(c)) in the wavelength range accessible by our lasers (1480-1680 nm). We note that the field symmetry indicates that this is a dark mode of the structure and thus should not couple directly to free space. However, due to the finite size of our membrane, we are able to break the structural symmetry to access the mode.[27] The optical resonance frequency can be tuned by controlling the membrane-substrate separation *s* (red line in Figure 1(d). Using numerical modeling performed in COMSOL Multiphysics, we find that the $g_{OM}$ (blue line, Figure 1(d)) of the mode is $2\pi \times -3.3$ GHz/nm at s = 350 nm and increases to $2\pi \times -150$ GHz/nm at s = 50 nm.

We explore this range of separations using novel techniques that we developed to leverage built-in stresses in the substrate. Commercial SOI wafers often have high levels of built in stress introduced during the process that bonds the device layer and buried oxide layer to the handle substrate,[28] but the sign and precise strength of the stress are often unknown. Furthermore, devices made in our double-SOI platform experience an additional bending moment tangential to the $SiO_2$ surfaces exposed during the HFVE process that produces an upward force on all undercut structures (see Supplement). To control these effects, we first introduce "accordion-like" arrays of narrow beams (Fig. 1(a) inset) to each arm that relieve the compressive (tensile) stress and prevent out-of plane buckling (breaking) through in-plane deformation of the accordion structure.[28] Next, we introduce shaped arrays of etch-holes at the base of each arm to modify the direction of the bending moment and thus control the resulting force on the membrane arms. Devices which do not contain either of these stress management techniques, such as the one pictured in Fig. 2(a), deflect upward by >350 nm, as shown in the optical profilometer measurement in Fig. 2(b). Introduction of the accordion-like structure and a rectangle-like array of etch-holes decrease the membrane deflection by over an order of magnitude (Fig. 2(d,e)). Importantly, by replacing the rectangle-like etch-hole pattern with a triangular one, we are able to deflect the membrane downward as seen in Fig. 2(g,h) by 105 nm. These experimental observations were verified by numerical modeling of the devices under the same compressive stress conditions performed in COMSOL Multiphysics (Fig. 2(c,f,i)).This is an important and novel feature of our platform that gives us independent lithographic control of the membrane-substrate separation of different devices on the same chip. We have fabricated devices (red circles, Fig. 1(d)) with separations as small as 135 nm, corresponding to a $g_{OM}$ of $-2\pi \times 65$ GHz/nm, which is the largest yet value of $g_{OM}$ seen in a repulsive system[20] by a factor of four. Finally, we note that our devices are designed such that the majority of device deflection occurs in the support arms. Simulations show that the membrane has a radius of curvature of 2 cm for 100 nm deflections, such that it remains essentially flat during our experiments. However, confocal measurements using an optical profilometer on fabricated devices indicate that anisotropy intrinsic to the HFVE process can induce tilts in our devices that can be as large as 10 nanometers from side to side. Further modifications to the device design and fabrication process should be able to minimize this tilt, making this system a good candidate not only for current MEMS technologies but also for other innovative structures requiring parallel-plate geometries, such as plate-plate Casimir oscillators.

Actuation in our system is achieved by a combination of optical and photothermal forces, the latter of which arises from absorption of light in the cavity that generates displacements through thermal expansion at the Silicon-SiO$_2$ interface.[6,15,16,29]. Light absorption can also directly change the optical properties of the mode by modifying the refractive index of the silicon through the thermal-optic effect (dn/dT). In previous studies, which featured wavelength-scale mode-volumes ($\sim (\lambda_0/n)^3$) and poor thermal transport properties,[30,31] the thermo-optic effect obscured the underlying optomechanics,[32] due in large part to the strong two-photon absorption in silicon at near-IR frequencies. By employing a system containing extended photonic crystal modes[26,33] with large optical mode volumes ($\sim$1500 $(\lambda_0/n)^3$) and thermal diffusion rates $\gamma_t$ ($\sim$ 450 kHz) orders of magnitude larger than those of microcavity geometries, we avoid the problems commonly associated with silicon optomechanical systems while still achieving optical forces comparable to those achieved with micocavities.

To quantify the photo-thermal-mechanical (PtM) effects relative to the optomechanical (OM) effects, we define coefficients $L_{OM} = \omega_0/g_{OM}$ and $L_{PtM} = [DC_{th}^{-1}\Gamma_{abs}]^{-1}(\Omega_m^2 + \gamma_t^2)/\gamma_t$ as the inverse of the optical and photo-thermal forces, respectively, per unit stored optical energy in the cavity, where D is the material- and geometry-dependent thermal-mechanical force coefficient in units of Newtons per Kelvin (positive or negative for attractive or repulsive forces, respectively), $C_{th}$ is the heat capacity, and $\Gamma_{abs}$ is the total absorption coefficient of the material. In our devices, thermal expansion causes a downward deflection, effectively generating an attractive force ($F_{pth}$) between the membrane and the substrate, which puts the photo-thermal force in competition with the repulsive optical force. We note that $F_{pth}$ and $L_{PtM}$ are essentially independent of membrane-substrate separation,[34] while, as shown in Fig 1(c), $F_{opt}$ and $L_{OM}$ are highly sensitive to changes in *s*.

Coupled-mode theory[35] is used to study how the coupling of the thermal, optical and mechanical degrees of freedom affects a device's mechanical angular frequency, $\Omega_m$ and its linewidth, $\Gamma_m$. (For full derivation, see Eq. 1-19 in supplement.) Physically, an optical cavity with linewidth $\kappa$ in a free-standing membrane undergoing Brownian motion (with an RMS amplitude $\delta s$ at $\Omega_m$) is perturbed by the optical field or thermal gradient such that the resonant frequency shifts ($\omega_0' = \omega_0 + g_{OM}\delta s + \frac{d\omega}{dT}\delta T$), thus modulating the detuning $\Delta_0' = \omega_l - \omega_0'$ between an incident laser source at $\omega_l$ and the shifted cavity resonance. This modulation in $\Delta_0'$ further modulates the stored cavity energy $U_{ph}$, the forces $F_{opt}$ and $F_{pth}$ and hence the overall displacement $\delta s$, which in turn modulates the optical resonance frequency and forms a feedback loop. This feedback has both in-phase and quadrature (out-of-phase) components that result in a mechanical frequency perturbation $\Omega_m$ and a mechanical linewidth perturbation $\Gamma_m$, respectively, both of which have odd symmetry with respect to $\Delta_0'$.

Modifications to $\Omega_m$ can take two forms. The applied forces can act in the same direction of the vibrational restoring force, resulting in a stiffening ($\delta\Omega_m > 0$), or in the opposite direction, resulting in softening ($\delta\Omega_m < 0$). Similarly, modifications to $\Gamma_m$, known as "induced back-action[7]," describe the direction of energy flow between the mechanical mode and the optical or thermal field. A net energy flow into the mechanical mode decreases $\Gamma_m$ and corresponds to mechanical amplification, while a net energy flow out of the mechanical mode increases $\Gamma_m$ and corresponds to increased damping (i.e. mechanical cooling). The strengths and signs of the modifications to the mechanical motion due to PtM and OM effects depend on the magnitudes of the respective forces ($L_{OM}^{-1}, L_{PtM}^{-1}$) as well as the strength of the optomechanical coupling, $g_{OM}$. This results in OM perturbations which are independent of the sign of the force ($\propto g_{OM}/L_{OM}$) and PtM perturbations which depend on the signs of both the optical and photothermal forces ($\propto g_{OM}/L_{PtM}$).

To study these dynamics and characterize our devices, we use the setup illustrated in Figure 3. Briefly, an optical fiber mounted on a z-translation stage centered above the suspended membrane in the Rugar configuration[36] (detail in inset i.) is used to couple light into a device and also collect the signal back-reflected by the device. The chip containing the device rests on a four-axis ($x,y,\theta_x,\theta_y$) stage platform. The motorized x-y stage contains closed-loop feedback that allows us to repeatably align the fiber to the sample with sub-100 nm resolution, while the manual $\theta_x$ and $\theta_y$ tilt stages are used to planarize the fiber facet to the chip surface. The whole setup is placed inside a high vacuum chamber ($10^{-5}$ torr) to eliminate gas damping of the mechanical vibrations. Optical spectra (Fig. 3, inset ii) were collected by sweeping the tunable laser sources across the optical resonance. Mechanical spectra (Fig. 3, inset iii) were collected at fixed excitation wavelengths on either side of the optical resonance by analyzing the signal reflected off of our membranes in a real-time spectrum analyzer (Tektronix RSA3303B). We measure the dynamic shifts in $\Omega_m$ and $\Gamma_m$ of the fundamental vibrational mode of two different devices by fitting the measured mechanical resonance to a Lorentzian lineshape, and the best-fit parameters for $\Omega_m$ and $\Gamma_m$ are plotted in Figure 4 (green circles) as a function of laser wavelength.

We first explore an upward-deflected device (Fig. 1(d)) with membrane-substrate separation s = 300 nm at an incident optical power of 50 µW, in Fig 4(a). We observe blue-detuned ($\Delta_0' > 0$) cooling and softening and red-detuned ($\Delta_0' < 0$) amplification and stiffening, which fit well to theoretical predictions (black lines) for $L_{OM}$ = -95 μm ($g_{OM}$ = 2π ×

-3.3 GHz/nm) and $L_{PtM}$ = 15 μm. As expected, PtM effects (red dashed lines) to dominate the

dynamics due to the large membrane-substrate separation. Furthermore, we see no OM

contributions (blue-dashed line) to the linewidth dynamics as a consequence of operating in the deep "sideband-unresolved" limit[7] where $κ/Ω_m$ > $10^6$. In this regime, the optical force acting on the mechanical oscillator is effectively instantaneous with respect to the oscillator period. All back-action in our devices is provided by PtM effects.

In a downward-deflected device with s = 160 nm, (Fig. 4(b), image in Fig. 2(e)), the

increased OM contributions (blue-dashed line) to the overall dynamics (black line) flips the

sign of the $δΩ_m$ curve compared to the upward-deflected device. The magnitudes of $δΓ_m$

and $δΩ_m$ are similar to those in Fig. 4(a) but were achieved with an order of magnitude less

optical power (6 μW). This is attributed to an increase in $g_{OM}$ to 2π × -30 GHz/nm ($L_{OM}$ = -

6.4 μm), while $L_{PtM}$ remained constant. When the incident optical power is increased to 30

μW (Figure 5(a)), a range of wavelengths (grey shaded region) exists in which $Γ_m$ reaches its experimental minimum.[37] In this region energy is being added to the mode faster than it can dissipate, resulting in a "regenerative oscillation" amplitude on the order of a nanometer that scales linearly with power. This regime is of interest for applications in mass-sensing. For example, the mechanical response at λ = 1561.5 nm (red star), plotted in optimization of the structure, with this application in mind, may result in even better mass sensitivity.

Fig 5(b), has $Γ_m/2π$ = 70 mHz, which corresponds to a mass sensing limit of 0.3 fg. Further

On the other side of the optical resonance, optical cooling takes place. The strength of the cooling is quantified by the mode's effective temperature $T_{eff}$,[6] which, from the equipartition theorem, we can write as $T_{eff}/T_0 = \Omega_{m0}^2/\Omega_m^2 \times \Gamma_{m0}/\Gamma_m$, where $\Gamma_m = \Gamma_{m0} + \delta\Gamma_m = \Gamma_{m0}(1 + \beta P)$, $T_0$ is room temperature and $\beta$ is a collection of constants defined precisely in the Supplementary Materials. We quantify the strength of the mechanical cooling by exciting the structure at a fixed wavelength (λ = 1560.8 nm, blue star in Fig. 5(a)) and at six optical powers in the range of 6 µW (red line) to 200 µW (purple line). We note that $T_{eff}$ is also proportional to the area under the mechanical resonance curves (blue shaded regions). As other processes can cause linewidth broadening without cooling, it is important to perform this check to confirm that we are indeed cooling the vibrational mode. The effective temperature of the mode at the six measured powers is shown in the Fig. 5(c) inset, where the colors of the circles correspond to the colors of the curves in the main figure. We find that $T_{eff}$ reaches 22 K at 200 µW and 5.8 K at 1 mW, according to the curve fit (black line). These values, coupled with the large device mass, large $g_{OM}$, and ease of in- and out-coupling of light, make this platform an intriguing candidate for optical accelerometry,[13] which utilizes optomechanical cooling to damp mechanical ringdown.

In summary, we have demonstrated a novel optomechanical platform based on a silicon photonic-crystal membrane suspended above an SOI substrate that exhibits a strong repulsive optical force. Using simple lithographic stress management techniques, we were able to control the membrane-substrate separation of our structures as well as their resulting optical and mechanical properties (e.g. resonance). The interplay between opto-mechanical and photo-thermal-mechanical effects was investigated, and both mechanical cooling and amplification have been demonstrated. Owing to its large size, our structure has many unique features including a large mass, ease of in- and out- coupling of light, and lack of two-photon absorption effects. Therefore, we believe that this structure is suitable for a range of applications. In addition to accelerometry and mass sensing mentioned above, our devices can function as large-area liquid or gas-phase sensors, sensitive to refractive index changes of the environment.[38] The antibonding mode studied here is uniquely suited for this application, as the electric field distribution is extremely sensitive to the refractive index around the holes in the suspended membrane (Fig. 1(c)). Preliminary simulations in COMSOL Multiphysics indicate that the spectral sensitivity of the cavity resonance is ≈ 100 nm per refractive index unit, on par with previous values seen in similar geometries.[39] With control over both the optical and mechanical degrees of freedom, our membranes are good candidates for selective sensing technologies, where differences in mass and optical properties of analytes can be distinguished from one another and where the trade-off between large surface area and mechanical sensitivity can be tolerated.[40,41]

Perhaps the most intriguing application of our platform is the investigation and control of the Casimir force. The Casimir effect causes two parallel surfaces separated by vacuum to be attracted to one another with a force proportional to $s^{-4}$ and can cause failure in MEMS and NEMS systems. High $Q_{opt}$, negative $g_{OM}$ modes offer a method to counterbalance the Casimir and electrostatic pull-in forces that cause stiction. Additionally, the Casimir force can profoundly modify the mechanical dynamics of the system[42,43] at separations of 100 nm or smaller, making this platform ideal for studying plate-plate Casimir dynamics. For example, the Casimir force acts as a nonlinear driving term to the mechanical oscillator, modifying $\Omega_m$ and introducing mechanical hysteresis, further improving the mass-sensing ability of the system by introducing bistability without the need for large oscillation amplitudes.[42] An integrated, tunable Casimir-mechanical oscillator is also desirable for testing of fundamental aspects of the Casimir effect, such as deviation from the proximity-force approximation[44,45] for an arbitrarily structured surface.

While our platform has many potential applications, we note that this system has not been optimized for any one specific application, and the estimates presented here should not be taken as the limits of this new geometry. Most of the device parameters ($Q_{opt}$, $Q_{mech}$, $g_{OM}$) can be improved by modifications to the design and fabrication process. For instance, annealing of some devices at 500 C for 1 hour in a nitrogen environment improved $Q_{opt}$ and $Q_{mech}$ by about a factor of two. Actuation and sensing abilities both improve as the membrane-substrate separation decreases, leading to the possibility of low-power devices that can take advantage of the large area of the suspended membrane, such as optical accelerometers and combined mass and refractive-index sensors that are enhanced by the Casimir effect. By building these devices on silicon, we have opened up a new pathway for integration of novel optical components into MEMS and NEMS systems.

**Acknowledgements**: This work was supported by the Defense Advanced Research Projects Agency (DARPA) under contract no N66001-09-1-2070-DOD. This work was carried out in part through the use of Massachusetts Institute of Technology's Microsystems Technology Laboratories, and in part at the Center for Nanoscale Systems (CNS) at Harvard University, a member of the National Nanotechnology Infrastructure Network (NNIN), which is supported by the National Science Foundation (NSF) under NSF award no. ECS-0335765. Thanks to Professor Rob Wood at Harvard University who owns the Olympus LEXT OLS4000 optical profilometer on which some of our measurements were performed. Additional thanks to Alexei Belyanin, Oskar Painter, Daniel Ramos, Igor Lovchinsky and Romain Blanchard for helpful discussions.

**Author Contributions:** Experiments and analysis were carried out by D.N.W and P.-C. H., with assistance from P.D.. Theoretical modeling was conducted by D.N.W, P.-C. H., A.W.R., and I.B.. M.K. performed the wafer bonding, and E.I. designed, fabricated, and characterized the devices, with input and feedback from D.N.W. and P.-C.H.. M.L., F.C., and S.G.J. oversaw the project. The paper was written by D.N.W. with input from all co-authors.

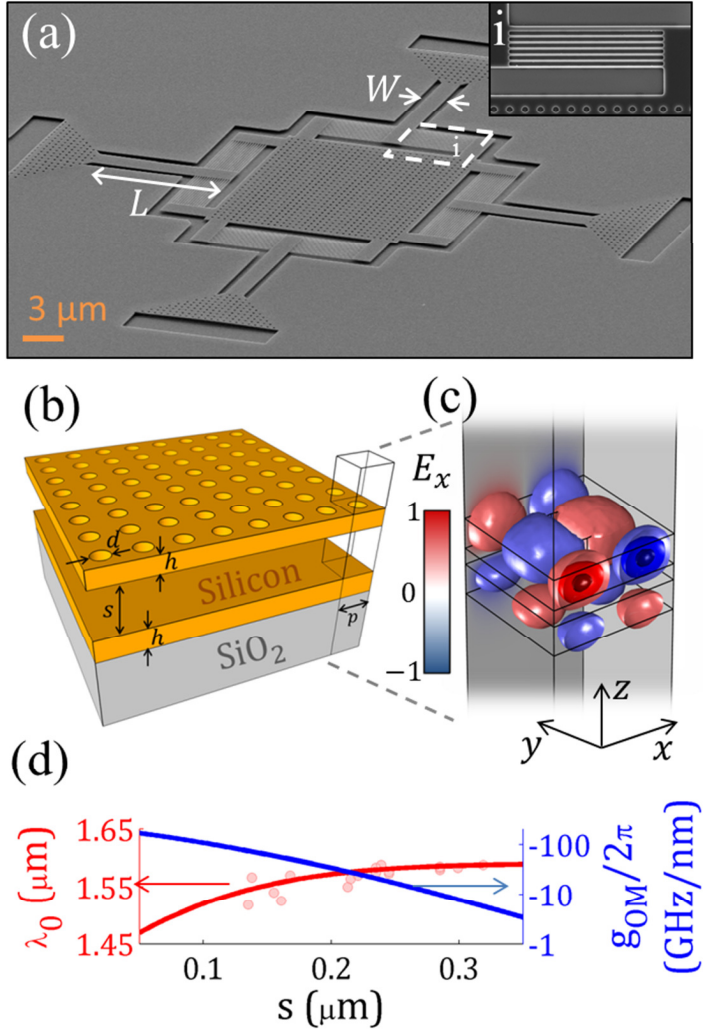

**Figure 1:** (a) SEM image of a device consisting of a h = 185 nm thick silicon membrane patterned with a square-lattice photonic crystal with a 30×30 periodic hole array of period p = 0.92 μm and hole diameter d = 0.414 μm suspended 165 nm above a Silicon-on-Insulator (SOI) substrate (h = 185 nm, buried oxide layer = 2 μm, cross-section shown schematically in (b)). The membrane is supported by four arms (L = 19.3 μm, W = 2.75 μm) that are terminated on their far ends by arrays of etch holes and on their near ends by "accordion-like" structures (inset i) which provide lithographic control of membrane-substrate separation. (c) A 3D optical mode simulation shows the x-component of the electric field of a single unit cell of the geometry in (b) with s = 100 nm for an antibonding mode of the structure at $\lambda_0$=1570 nm. The antisymmetric field symmetry with respect to the gap between the membrane and the substrate implies that this optical mode generates a repulsive force. (d) The calculated resonance wavelength $\lambda_0$ (red line) of the mode in (c) is plotted with data points (red circles) representing 16 different devices with identical membrane designs but different membrane-substrate separations. The separations were determined by interferometric measurements using an optical profilometer. The blue line is optomechanical coupling coefficient of the mode and is proportional to the slope of the red line ($g_{0M} \propto -d\lambda/ds$).

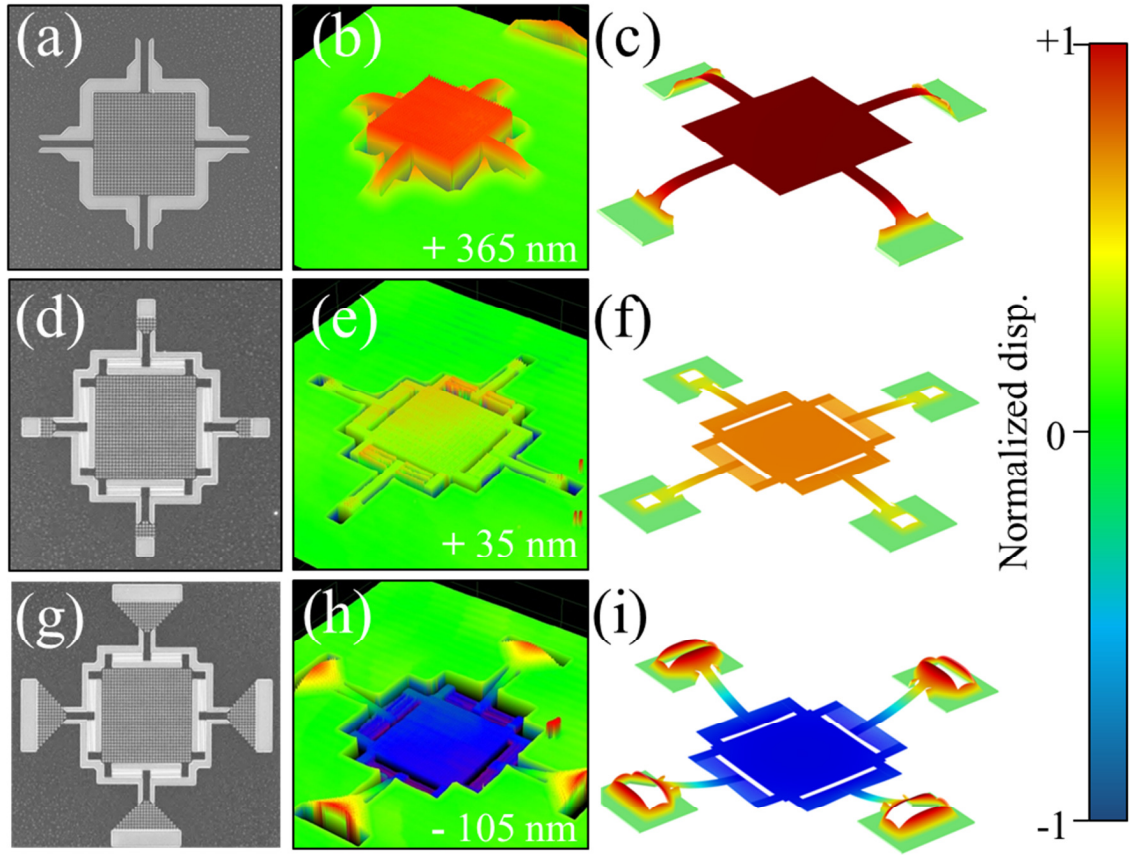

**Figure 2:** Scanning Electron Microscope (SEM) images (a,d,g), optical profilometer images showing interferometric measurements of height (b,c,h) and mechanical simulations in COMSOL Multiphysics confirming the behavior measured by the profilometer (d,f,i) are shown for three separate devices. The first device (a) has simple support arms which offer no control over the stresses inherent in the wafer. The result is a membrane which deflects strongly upward by 365 nm (b,c) due to compressive stress in the membrane and an upward torque caused by a bending moment at the Si-SiO$_2$ etch boundary. The second device (b) has an accordion-like structures between the arms and the membrane (see inset, Fig. 1(a)) and an approximately rectangular etch-hole pattern at the base of each arm to combat compressive stress and the bending moment at the etch boundary. The result is a device which only deflects upward 35 nm (e,f), an order of magnitude improvement over the device in (a). The triangular design of the etch-hole pattern in the third device (g) utilizes the torque induced at the etch boundary to generate a controllable downward deflection. The result (h,i) is a membrane which deflects downward 105 nm from the surrounding silicon layer.

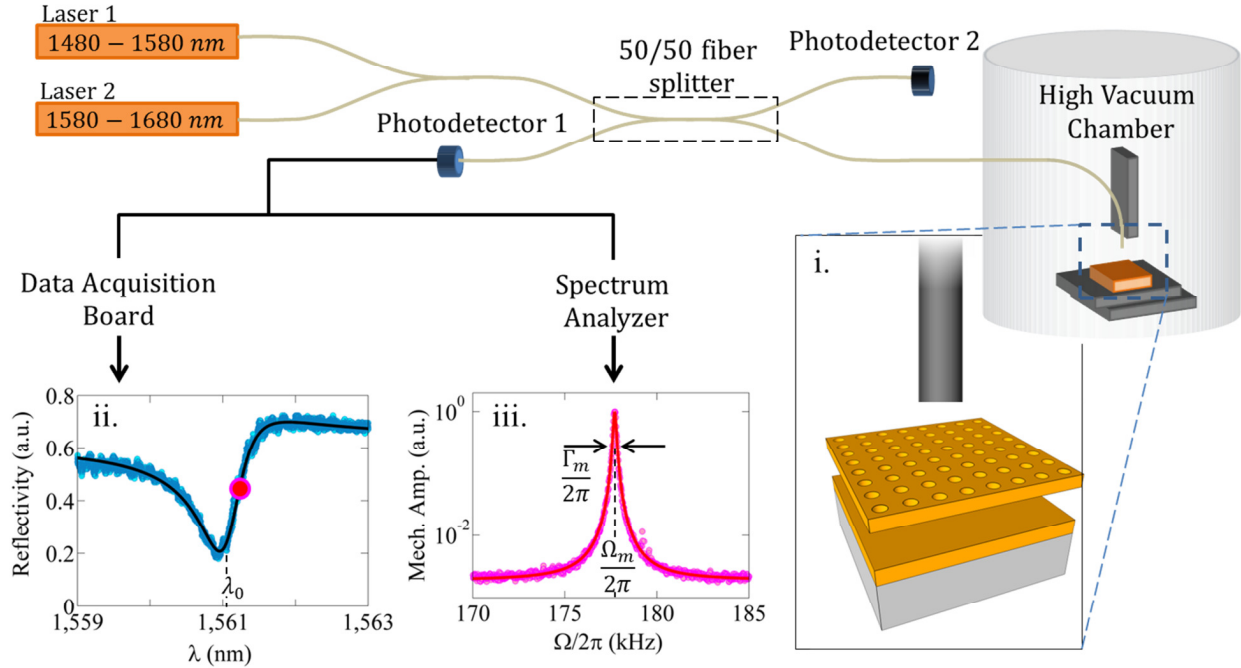

**Figure 3:** Experimental apparatus. The outputs from two tunable near IR lasers ($\lambda = 1480$ nm $-$ 1580 nm and $\lambda = 1580$ nm $- 1680$ nm) are combined using a fiber-directional coupler. Half of the signal is diverted to Photodetector 2 as a reference and half is coupled into a high –vacuum chamber (HVC) via a custom made fiber-feedthrough port. The cleaved fiber is positioned above the center of the device, so that the cleaved fiber facet is parallel to the membrane (inset i., not to scale). The reflected optical signal is measured at Photodetector 1. Optical reflection spectra are taken by sweeping the lasers' wavelengths across the optical resonances and collecting the signal via the Data Acquisition (DAQ) board. The optical resonance centered at $\lambda_0$ = 1561.1 nm (inset ii) has a Fano shape (black line) and $Q_{opt}$ = 2500. Mechanical spectra are obtained by taking the Fourier transform of the photodetector signal using the spectrum analyzer (SA) to measure the small thermal vibrations of the membrane. The fundamental mechanical resonance (inset iii), defined by resonance frequency $\Omega_m$ and linewidth $\Gamma_m$, is shown for a low-power measurement at $\lambda$=1561.2 nm (red dot, inset ii.).

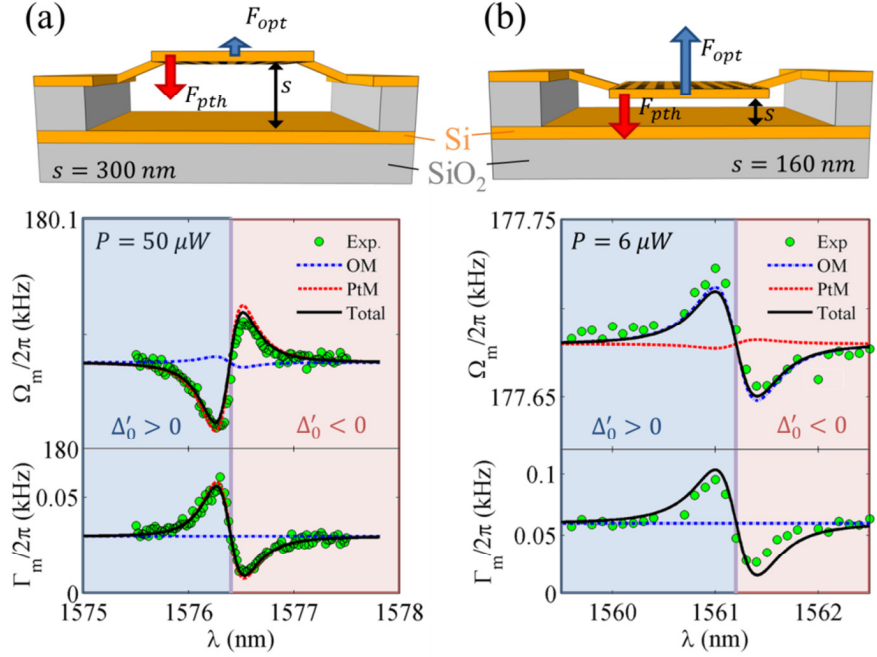

**Figure 4:** Optomechanical coupling curves for two devices with different membrane-substrate separations experiencing photothermal and optical forces. Data (green circles) were collected by measuring the vibrational spectra (Fig. 3, inset iii.) for several wavelengths around the cavity resonance. The data in (a) correspond to the device in Fig. 2(d), whose cavity resonance is centered at $\lambda_0$=1576.4 nm. The data in (b) correspond to the device in Fig. 2(g), whose cavity resonance is centered at $\lambda_0$=1561.1nm. The photothermal force $F_{pth}$ (red arrows, top insets) is attractive and approximately constant in both devices. The repulsive optical force $F_{opt}$ (blue arrows), increases in magnitude from (a) to (b), as the magnitude of $g_{OM}/2\pi$ increases from -3.3 GHz/nm to -30 GHz/nm. In our system, $F_{opt}$ and $F_{pth}$ have opposing effects on $\Omega_m$ (top panels). In (a), photo-thermal-mechanical (PtM) dynamics (red dashed lines) dominate and the device undergoes softening ($\delta\Omega_m < 0$) when the laser is blue-detuned ($\Delta'_0 > 0$, blue shaded region) and stiffening ($\delta\Omega_m > 0$) when red-detuned ($\Delta'_0 < 0$, red shaded region). In (b), optomechanical (OM) dynamics (blue dashed lines) dominate and the device undergoes blue-detuned stiffening and red-detuned softening. Bottom panels: Both devices undergo blue-detuned cooling ($\delta\Gamma_m > 0$) and red-detuned amplification ($\delta\Gamma_m < 0$) due to PtM effects only. The maximum values of $\delta\Omega_m$ and $\delta\Gamma_m$ shown in both (a) and (b) are approximately equal in magnitude, but the dynamics in (b) were achieved with an order of magnitude less optical power due to the of strength of opto-mechanical coupling ($g_{OM}$) in (b).

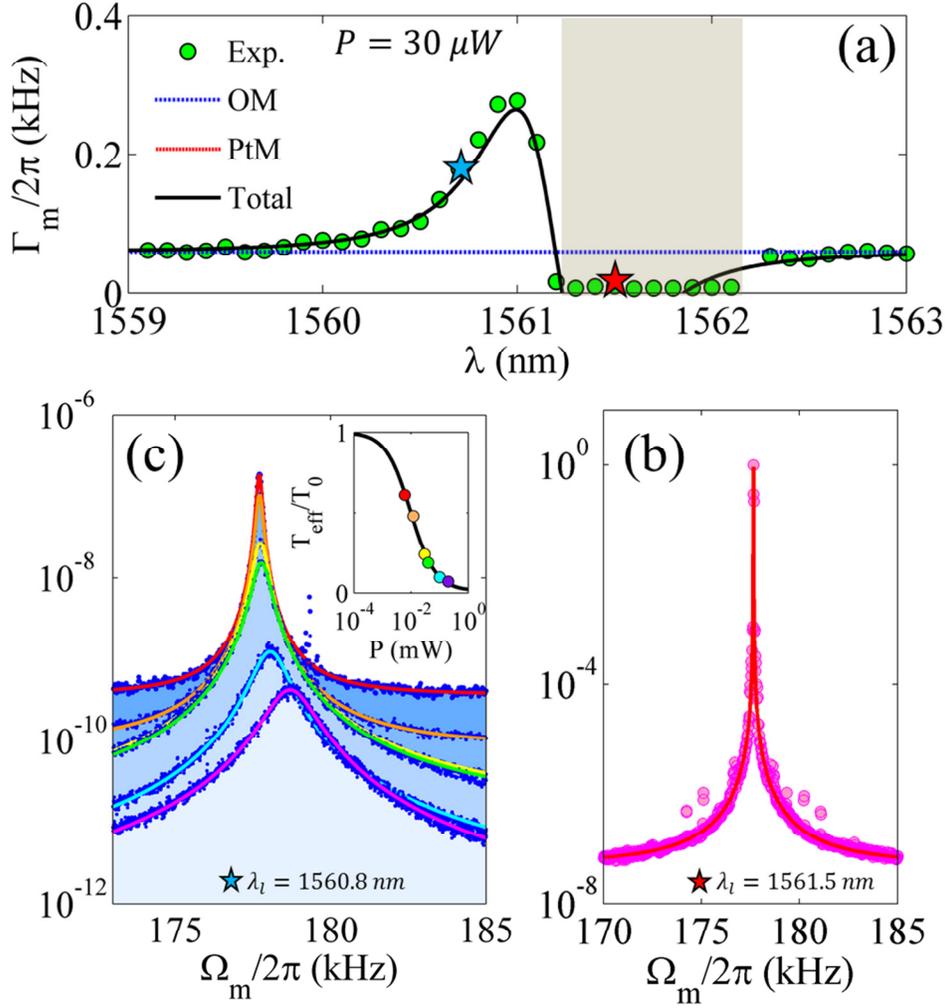

**Figure 5:** (a): Mechanical linewidth of the device (green circles) in Fig 4(b) as a function of laser wavelength at an incident power of 30 µW, showing blue-detuned cooling and red-detuned amplification. The overall dynamics (black line) are dominated by photo-thermal mechanics (red dashed line, not seen), since optomechanical interactions (blue dashed line) are negligible. On the red side of the resonance between 1561.2 nm and 1562.1 nm (grey shaded region of (a)), the mechanical linewidth hits a floor as the system undergoes generative oscillations. (b): The vibrational spectrum of the mode when $\lambda_l$ = 1561.5 nm is shown (pink circles) with the lorentzian fit (red line), where $\Gamma_m/2\pi$ = 70 mHz. (c): On the other side of the resonance, the mechanical vibration is cooled. The mechanical resonance is plotted (dark blue dots) and fit for six powers: 6 (red line), 12 (orange line), 30 (yellow line), 40 (green line), 100 (blue line), and 200 µW (purple line). In our system, linewidth broadening is due to mechanical cooling only. Thus, $\Gamma_m$ and the area under the mechanical resonance curves (blue shaded regions) are both be proportional to the effective temperature of the mode ($T_{eff}$), which is plotted as a function of power in the inset. The colors of the data points correspond to the colors of the lines in (c), and the points are fit to the expression in the text. At 1 mW, the effective temperature of the mode is 5.6 K when cooled from room temperature.

**METHODS**

**Fabrication** First, two high resistivity Silicon-on Insulator (SOI) wafers (SOITEC, device layer thickness = 220 nm, Buried Oxide (BOx) thickness = 2 µm) were thinned down using thermal oxidation, reducing the silicon device layer thickness to 185 nm. Next, oxide-oxide bonding was used to bind the two wafers together. After low-stress nitride passivation, we then dry etch away the backside of one of the handle silicon wafers using KOH to wet-etch the exposed handle silicon. This is followed by a Buffered Oxide (7:1 $H_2O$:HF) etch to remove the exposed box layer to reveal two 185 nm silicon device layers separated by 260 nm of thermal oxide, above the 2 µm BOx layer. Devices patterns were then written using conventional 100 keV e-beam lithography on ZEP-520 positive photoresist and transferred to the top device layer using inductively-coupled plasma reactive-ion etching. Finally, the oxide layer between the two Si slabs is removed by Hydrogen Fluoride Vapor-phase Etching to release the top device layer and provide the gap between the top and bottom membrane. The process is described visually in the supplement (Figure S2(a)). Annealing of some devices (data not shown) was performed at 500 C for 1 hour in a nitrogen environment in order to maximize optical and mechanical quality factors.

**Measurement** Measurements were made after parallelizing the fiber facet to the substrate from fiber-substrate heights of approximately 25 µm (Fig. 3, inset i) corresponding to a spot size of approximately 12 µm. This configuration allows straightforward coupling into the evanescently-coupled lateral modes of the PC membrane-substrate geometry at the Γ point of the band structure. In all experiments, we chose the fiber-membrane alignment which maximized coupling.

# SUPPLIMENTARY MATERIAL

1. **Derivation of Dynamics**

Cavity optomechanical dynamics can be represented by a system of two equations describing the stored cavity energy and the mechanical motion:

(1) $$\frac{da}{dt} = -\frac{\kappa}{2}a - i(\Delta + g_{OM}x)a + \sqrt{\kappa_e}\,\epsilon$$

(2) $$\frac{d^2x}{dt^2} + \Gamma_m \frac{dx}{dt} + \Omega_m^2 x = -\frac{g_{OM}|a^2|}{m_{eff}\omega_0}$$

Equation 1 describes the properties of the optical cavity, where $a$ is the amplitude of the optical field in the mode, $\epsilon$ is the amplitude of the field incident on the structure at $\omega_l$, $\kappa_e$ is the external coupling rate such that $\kappa_e|\epsilon|^2$ is the incident optical power, and $\Delta = \omega_0 - \omega_l$. Equation 2 describes the mechanical behavior, where $x$ is the oscillation amplitude, $\Gamma_m$ and $\Omega_m$ are the mechanical linewidth and frequency, respectively, and $m_{eff}$ is the effective mass of the mechanical element. The perturbations to $\Omega_m$ and $\Gamma_m$ from opto-mechanical coupling arise from in-phase ($\propto x(t)$) and quadrature ($\propto \dot{x}(t)$) driving terms driving the harmonic oscillator, and can be obtained by linearizing (1) and (2). These expressions can be written as functions of $\Delta$ in the so-called "sideband-unresolved" limit[7] ($\Omega_m \ll \kappa/2$) as

(3) $$\delta\Omega_m = \Omega_m \frac{g_{OM}^2}{\omega_0^2 K_m}\left(\frac{\frac{\kappa_e}{2}|\epsilon|^2}{\frac{\kappa^2}{2} + \Delta_0^2}\right)\frac{2\Delta\omega_o}{\frac{\kappa^2}{2} + \Delta_0^2}$$

and

(4) $$\delta\Gamma = -\Omega_m \frac{g_{OM}^2}{\omega_0^2 K_m}\left(\frac{\frac{\kappa_e}{2}|\epsilon|^2}{\frac{\kappa^2}{2} + \Delta_0^2}\right)\frac{2\Delta\omega_o}{\left(\frac{\kappa^2}{2} + \Delta_0^2\right)}\frac{2\Omega_m \frac{\kappa}{2}}{\left(\frac{\kappa^2}{2} + \Delta_0^2\right)}$$

where $K_m = \Omega_m^2 m_{eff}$ is the mechanical spring constant, and $\Delta_0 = \Delta - g_{OM}x$. In this limit, it is clear from the above equations that the change in the linewidth of the resonator must be significantly smaller than the change in the mechanical frequency. The two equations differ only by the term at the far right of (4), relating $\Omega_m$ to $\kappa/2$, which by definition is $\ll 1$ in this limit.

When thermal-optical and thermal-mechanical effects are included, equations (1) and (2) must be modified and a third equation describing the system's thermal properties must be added. The system then becomes:

(5) $$\frac{da}{dt} = -\frac{\kappa}{2}a - i\left(\Delta - g_{OM}x - \frac{d\omega_0}{dT}\delta T\right)a + \sqrt{\kappa_e}$$

(6) $$\frac{d^2x}{dt^2} + \Gamma_m \frac{dx}{dt} + \Omega_m^2 x = -\frac{g_{OM}|a|^2}{m_{eff}\omega_0} - \frac{D}{m_{eff}}T$$

(7) $$\frac{dT}{dt} = -\gamma_{th}T + C_{th}^{-1}(\Gamma_{lin} + \Gamma'_{TPA}|a|^2)|a|^2$$

where $D$ is the thermal mechanical force coefficient in units of $N/K$, ($D > 0$ for downward thermally-induced displacements), $\gamma_{th}$ is the thermal time constant, $C_{th}$ is the thermal heat capacity, $\Gamma_{lin}$ is the linear absorption rate of silicon and $\Gamma'_{TPA}$ is the two-photon absorption rate (included for competition) with the dependence on field intensity $|a|^2$ explicitly removed from the expression for clarity, such that the total absorption rate is $\Gamma_{abs} = \Gamma_{lin} + \Gamma'_{TPA}|a|^2$. To solve this system for its mechanical dynamics, we must first linearize it[5], rewriting $x(t) = x_0 + \delta x(t)$, $a(t) = a_0 + \delta a(t)$, and $T(t) = T_0 + \delta T(t)$. The linearization of Eq. 7 results in separate expressions for $T_0$ and for $\delta T(t)$, though for this derivation, we are only interested in the equation for $\delta T(t)$, which takes the form

(8)
$$\frac{d}{dt}\delta T(t) = -\gamma_{th}\delta T(t) + c_{th}^{-1}\Gamma_{abs}\big(a_0\delta a^*(t) + a_0^*\delta a(t)\big).$$

Converting into Fourier space, this expression becomes

(9)
$$\delta T(\omega) = \frac{c_{th}^{-1}\Gamma_{abs}}{-i\omega + \gamma_{th}}(a_0\delta a^* + a_0^*\delta a),$$

where the expression for $a_0\delta a^* + a_0^*\delta a$ can be found by linearizing Eq. 5 and solving for $a_0$ and $\delta a$ independently, then converting to Fourier space. Doing this, one finds

(10)
$$a_0\delta a^* + a_0^*\delta a = i|a_0|^2\left(g_{OM}\delta x(\omega) + \frac{d\omega}{dT}\delta T(\omega)\right)\left[\frac{1}{\frac{\Gamma}{2} + i(\omega + \Delta_0')} - \frac{1}{\frac{\Gamma}{2} + i(\omega - \Delta_0')}\right].$$

Plugging (9) into (10) we find that

(11)
$$a_0\delta a^* + a_0^*\delta a = i|a_0|^2 g_{OM}\delta x(\omega)H(\omega,\Delta_0',|a_0|^2)$$

where

(12)
$$H(\omega,\Delta_0',|a_0|^2) = \left(\left[\frac{1}{\frac{\Gamma}{2} + i(\omega + \Delta_0')} - \frac{1}{\frac{\Gamma}{2} + i(\omega - \Delta_0')}\right]^{-1} - i|a_0|^2\frac{d\omega}{dT}\frac{c_{th}^{-1}\Gamma_{abs}}{-i\omega + \gamma_{th}}\right)^{-1}.$$

Ultimately we are interested in the perturbations to $\Omega_m$ and $\Gamma_m$, which will need to be expressed as in-phase and quadrature driving terms in (6). In order to show this, we now force $\delta x(t)$ and $\delta a(t)$ to be harmonically oscillating functions at $\Omega_m$, allowing us to write $\delta x(\omega)$ as $\frac{1}{2}\big(\delta(\omega - \Omega_m) + \delta(\omega + \Omega_m)\big)$. Using this and separating the real and imaginary parts of $H(\Omega_m)$, we can rewrite (11) as

(13)
$$a_0\delta a^* + a_0^*\delta a = \frac{1}{2}g_{OM}|a_0|^2(i\Re[H(\Omega_m)]\big(\delta(\omega-\Omega_m)-\delta(\omega+\Omega_m)\big)$$
$$-\Im[H(\Omega_m)]\big(\delta(\omega-\Omega_m)+\delta(\omega+\Omega_m)\big)),$$

which in the time domain becomes

(14)
$$a_0\delta a^*(t) + a_0^*\delta a(t) = g_{OM}|a_0|^2\left(\frac{\Re[H(\Omega_m)]}{\Omega_m}\delta\dot{x}(t) - \Im[H(\Omega_m)]\delta x(t)\right).$$

From here we plug (13) back into (9) and find

(15)
$$\delta T(\omega) = -\frac{c_{th}^{-1}\Gamma_{abs}|a_0|^2 g_{OM}}{\Omega_m^2+\gamma_{th}^2}\bigg[(\gamma_{th}\Re[H(\Omega_m)]$$
$$-\Omega_m\Im[H(\Omega_m)])\frac{\delta(\omega-\Omega_m)-\delta(\omega+\Omega_m)}{2i}$$
$$-(\Omega_m\Re[H(\Omega_m)]+\gamma_{th}\Im[H(\Omega_m)])\frac{\delta(\omega-\Omega_m)+\delta(\omega+\Omega_m)}{2}\bigg],$$

which in the time domain is

(16)
$$\delta T(\omega) = -\frac{c_{th}^{-1}\Gamma_{abs}|a_0|^2 g_{OM}}{\Omega_m^2+\gamma_{th}^2}\bigg[\delta\dot{x}(t)\frac{\gamma_{th}\Re[H(\Omega_m)]-\Omega_m\Im[H(\Omega_m)]}{\Omega_m}$$
$$+\delta x(t)\,(\Omega_m\Re[H(\Omega_m)]+\gamma_{th}\Im[H(\Omega_m)])\bigg].$$

We can now go back to our original system of equations and linearize (6), finding

(17)
$$\frac{d^2}{dt^2}\delta x(t) + \Gamma_m\frac{d}{dt}\delta x(t) + \Omega_m^2\delta x(t)$$
$$= -\frac{g_{OM}}{m_{eff}\omega_0}(a_0\delta a^*(t)+a_0^*\delta a(t)) + \frac{D}{m_{eff}}\delta T(t)$$

After plugging (14) and (16) into the RHS of (17) it becomes clear that (17) can be rewritten as a simple driven harmonic oscillator with perturbations to $\Omega_m$ and $\Gamma_m$ by grouping the terms on the RHS that contain $\delta\dot{x}$ (the quadrature terms) with $\Gamma_m$ and the terms that contain $\delta x$ (the in-phase terms) with $\Omega_m^2$. Doing this, we find that the system of equations involving thermal, optical, and mechanical dynamics results in mechanical perturbations of the forms

$$(18) \quad \frac{\delta\Omega_m}{\Omega_m} = -\frac{|a|^2}{2K_m}\frac{g_{OM}}{L_{OM}} Im[H(\Omega_m)] - \frac{|a|^2}{2K_m}\frac{g_{OM}}{L_{PtM}}\left(\frac{\Omega_m}{\gamma_{th}}Re[H(\Omega_m)] + Im[H(\Omega_m)]\right)$$

$$(19) \quad \frac{\delta\Gamma_m}{\Omega_m} = \frac{|a|^2}{K_m}\frac{g_{OM}}{L_{OM}} Re[H(\Omega_m)] + \frac{|a|^2}{K_m}\frac{g_{OM}}{L_{PtM}}\left(Re[H(\Omega_m)] - \frac{\Omega_m}{\gamma_{th}}Im[H(\Omega_m)]\right)$$

We define $L_{PtM} \equiv [DC_{th}^{-1}\Gamma_{abs}]^{-1}(\Omega_m^2 + \gamma_t^2)/\gamma_t$ and $L_{OM} \equiv \omega_0/g_{OM}$, which can be physically understood as the inverse of the photo-thermal and optical forces per unit stored optical energy, respectively. From linearizing (5), we find that the circulating optical field can be expressed in terms of the incident power as $|a|^2 = \frac{\frac{\kappa_e}{2}|\epsilon|^2}{\frac{\kappa^2}{2} + \Delta_0'^2}$. Eqs. 18 and 19 are used to fit the data in shown in the main paper and the curves in Figure S1. We plot in Fig. S1 the $\Omega_m$ and $\Gamma_m$ dynamics for the deflected down device at four different optical powers: 6 $\mu W$ (red circles), 12 $\mu W$ (orange circles), 20 $\mu W$ (yellow circles), and 100 $\mu W$ (blue circles). The best fit parameters reveal a photo-thermal force which scales linearly with power, and thus corresponds to linear absorption. However, the absorption rate $\Gamma_{abs}$ seems to be larger than that expected from bulk silicon, most likely due to absorption from surface defects and surface adsorbents introduced during the fabrication process[46,47], though it is difficult to decouple $\Gamma_{abs}$ from D and thus know $\Gamma_{abs}$ precisely. Measuring a set of devices before and after an annealing process seemed to confirm this, as both the mechanical quality factor and optical quality factor improved slightly after baking. We can find an upper bound on $\Gamma_{abs}$ by finding the rate at which the thermal-optic effect (last term in (12)) becomes notable, as our data reveals the effect to be quite small in our system. From this, we can estimate a value of the absorption quality factor: ($Q_{abs} = \omega_0/\Gamma_{abs} \approx 10^5$).

### 2. Controllable deflections via stress engineering

The fabrication procedure, described in the methods section, is illustrated in Figure S2(a). A cross section of the layer stack taken using a Scanning Electron Microscope is shown in Fig. S2(b). The whole process of thermal oxidation, oxide-oxide bonding, and removal of the handle silicon results in stresses in the multi-thin film layer structure, which leads to buckling of the silicon device layer when it is released. As a consequence of these processing and fabrication steps, we have also observed faster oxide HFVE rates in the bond-interface region that result in the presence of residual oxide close to the silicon membrane, which can be wider than 1 µm at the edges of the etch area (Figure S1(c)). A bending moment $M_\parallel$ due to the residual stress in the device layer (curved purple lines, Figure S3(a)i) causes a strong upward force on the devices fabricated with simple arms (Fig. 2(a) and S3(b)i), forcing them to deflect more than 300 nm.

In order to control this effect and engineer the direction of the bending moment, we devised an approach that rotates the axis of the bending moment by 90 degrees ($M_\perp$) by etching a small array of holes at the base of each arm and a large trench behind the etch-hole array. A simple rectangular etch-hole pattern (Figure S3(a) ii.) forms a "bridge" and forces the bending moment to act in a direction of much greater mechanical stiffness (red arrows), resulting in a significantly smaller upward deflection of the device (Fig. 2(d) and S3(b) ii.). If we instead use a triangular etch-hole array (Fig. S3(a) iii), we make the bridge less stiff farther from the device arm, resulting in a large upward deflection in the back of the bridge (large green arrow), while only generating a small deflection in the front (small green arrow). More importantly, however, is that this geometry generates a downward force on the device, whose strength is controllable by the pitch of the triangular array shape. Fig S3(c) and (d) show optical and profilometer images of structures made up of two device arms and two sets of accordion structures in the center. As is clear in the images, the devices with the widest triangle etch-hole pattern (i.) generate the largest downward deflection (blue color in (d)). The narrower triangle pattern (ii.) generates a small downward deflection, the rectangular pattern (iii.) results in a small upward deflection, and the etch-hole free device (iv.) suffers from a large upward deflection.

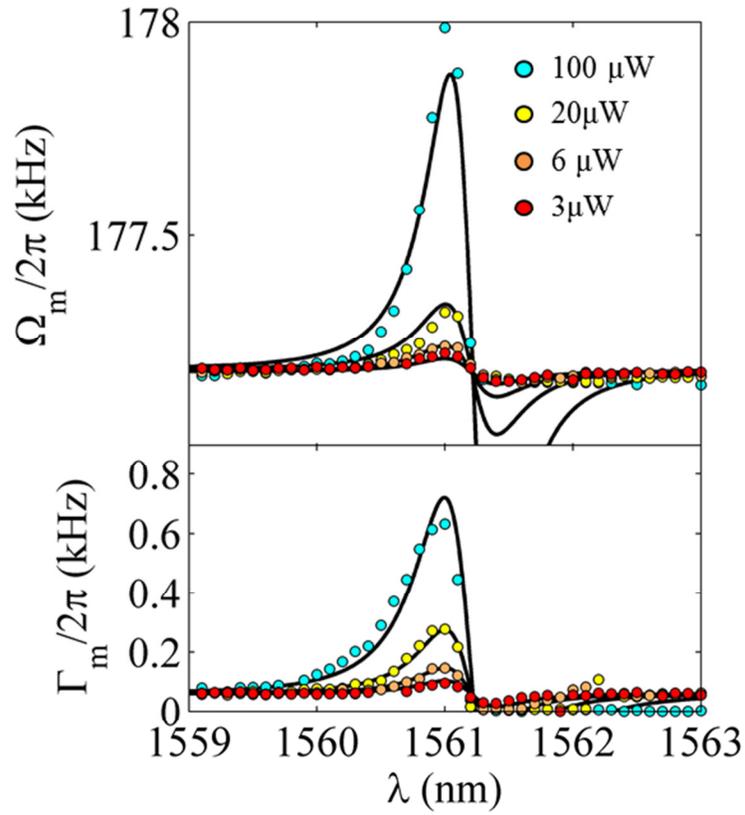

**Figure S1:** Mechanical frequency and linewidth at four different power levels (colors correspond to colors in Fig. 5(c)). The same fit parameters were used in all fitting curves, demonstrating a quality fit across multiple data sets and powers on the same device. Note that the fit to the photo-thermal-mechanical dynamics has linear power dependence, seen most clearly in the $\Gamma_m$ dynamics (bottom panel) thus demonstrating the lack of two-photon absorption, which would have quadratic dependence.

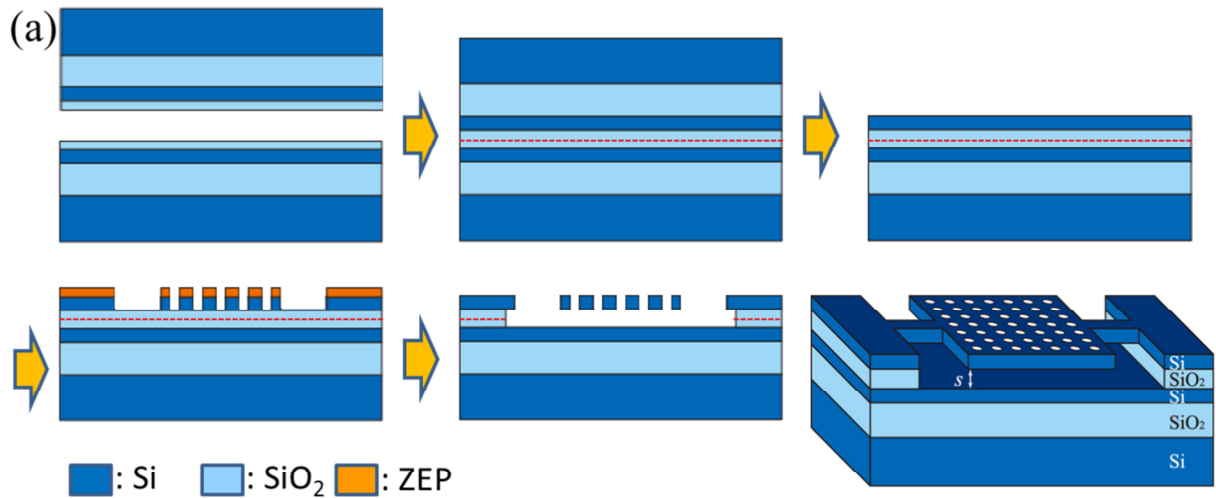
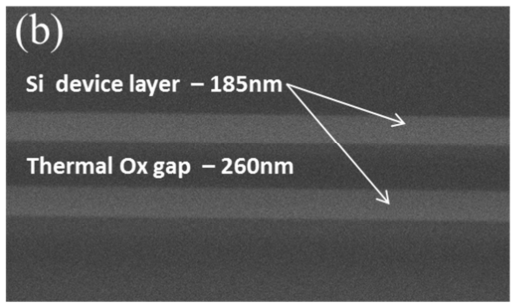
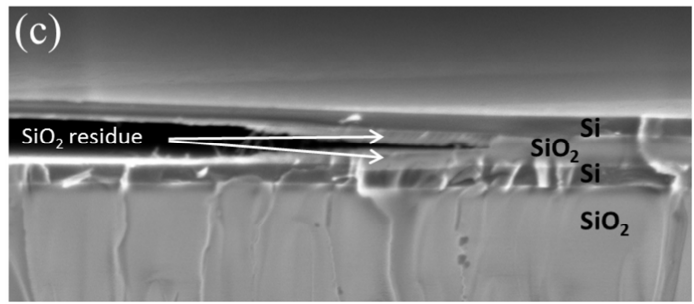

**Figure S2:** (a): Fabrication process. Two SOI wafers are bonded together with an oxide-oxide bonding process, creating a sandwich structure with two thin silicon device layers in the middle. SEM image of the sandwich shown in (b). Removal of one of the handle wafers with KOH and the thick oxide layer with BOE leaves a double-device layer chip, with the two Si layers separated by a 260 nm oxide gap. E-beam lithography followed by Reactive Ion Etching of the top silicon layer and HF Vapor Etching of the thermal-oxide gap layer create the final device (last panel of (a). (c): HFVE selectively etches at a higher rate along the oxide-oxide bond interface (red-dashed line) than it etches through the oxide layer, resulting in $SiO_2$ residue along the Silicon surfaces.

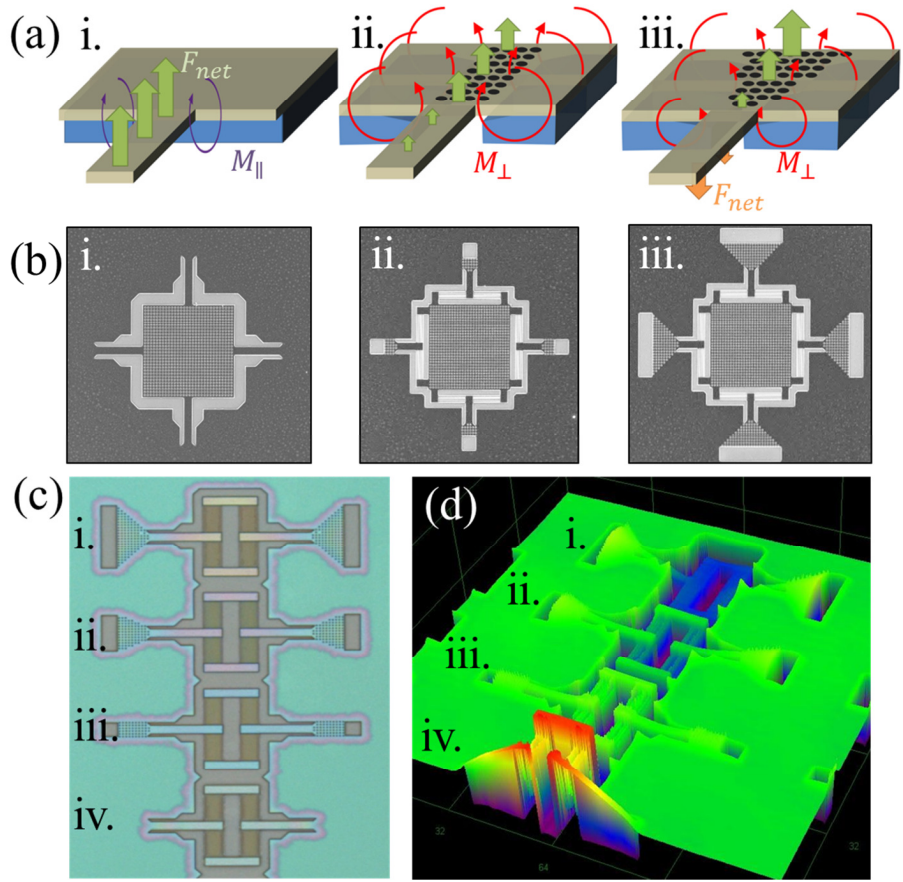

**Figure S3:** (a) Force diagram demonstrating our method for counterbalancing and counteracting the upward deflection caused by the oxide residue. i. With no modifications, the bending moment $M_\parallel$ (purple arrows) caused by the undercut $SiO_2$ (rendered in blue) etch-boundary produces a strong upward force (green arrows) on the membrane arms. ii. Etch-holes are introduced in a rectangular array at the base of the arm, forming "bridges" and changing the axis of the bending moment ($M_\perp$, red arrows). This results in an upward torque along the much stiffer axis which results in a smaller upward deflection of the device arms. iii. The shape of the hole array is triangular, making the bridge less stiff farther from the device arm, which results in a larger upward deflection in the back of the bridge than in the front of the bridge. The net effect of this tilt is to deflect the device arms downward (orange arrows). Devices corresponding to the diagrams in (a) are shown in (b), as well as Fig. 2 in the main text. Optical image (c) and interferometric measurement of the height profile (d) taken using the optical profilometer of four devices consisting of only two support beams and the "accordion" structure meant to absorb compressive/tensile stress. The termination of the arms varies and corresponds to the net deflection of the device, as follows: i. Wide triangle (large downward deflection) ii. Narrow triangle (small downward deflection) iii. Rectangle (small upward deflection) iv. No etch holes (large upward deflection).